\documentclass[letterpaper]{article} 
\usepackage[preprint]{aaai2027}  
\usepackage[hyphens]{url}  
\usepackage{graphicx} 
\usepackage{natbib}  
\usepackage{caption} 
\usepackage{algorithm}
\usepackage{algorithmic}

\usepackage{amsmath}
\usepackage{amssymb}
\usepackage{booktabs}
\usepackage{multirow}

\usepackage{newfloat}
\usepackage{listings}
\DeclareCaptionStyle{ruled}{labelfont=normalfont,labelsep=colon,strut=off} 
\floatstyle{ruled}
\newfloat{listing}{tb}{lst}{}
\floatname{listing}{Listing}

\usepackage{booktabs}

\title{RobustFlow: Towards Robust Agentic Workflow Generation}
\author{
    Shengxiang Xu\textsuperscript{\rm 1}\equalcontrib,
    Jiayi Zhang\textsuperscript{\rm 2}\equalcontrib,
    Shimin Di\textsuperscript{\rm 1}\corresponding,
    Yuyu Luo\textsuperscript{\rm 2}\corresponding,\\
    Liang Yao\textsuperscript{\rm 3},
    Hanmo Liu\textsuperscript{\rm 4,\rm 2},
    Jia Zhu\textsuperscript{\rm 5},
    Min-Ling Zhang\textsuperscript{\rm 1}
}

\affiliations{
    \textsuperscript{\rm 1}Southeast University;
    \textsuperscript{\rm 2}The Hong Kong University of Science and Technology (Guangzhou)\\
    \textsuperscript{\rm 3}Hohai University;
    \textsuperscript{\rm 4}Hong Kong University of Science and Technology;
    \textsuperscript{\rm 5}Zhejiang Normal University\\
    shimin.di@seu.edu.cn, yuyuluo@hkust-gz.edu.cn
}

\begin{document}

\maketitle

\begin{abstract}
The automated generation of agentic workflows is a promising frontier for enabling large language models (LLMs) to solve complex tasks. 
However, the empirical study
reveals that existing agentic workflow generation methods are not robust. 
They often generate inconsistent workflows when provided with instructions that are semantically equivalent but phrased differently. This brittleness severely undermines their reliability in real-world applications. 
To tackle this challenge, we propose RobustFlow, a robust agentic workflow generation system that leverages preference optimization to learn invariance across instruction variations. 
We also introduce a benchmark of semantically equivalent instruction variants with node-level and graph-level metrics for evaluating workflow generation robustness. 
By training on these instruction variants, RobustFlow achieves workflow generation robustness scores of 70\%--90\% across diverse perturbations, outperforming existing approaches including AFlow and ScoreFlow.
\end{abstract}

\section{Introduction}
The paradigm of evolving from independent large language models to orchestrated multi-agent systems has emerged as a highly promising direction for tackling complex, open-ended real-world tasks. By decomposing intricate and long-horizon problems into a structured sequence of collaborative actions, such agentic workflows can significantly enhance the planning and reasoning capabilities of LLMs~\citep{khattab2024dspy,tang2023verifai}. However, traditional multi-agent systems often rely on manually crafted workflows and static rules, which are labor-intensive and generalize poorly to unseen tasks~\citep{hong2024metagpt,qian2023chatdev}. Consequently, the automated generation of such workflows has become a critical area of research, aiming to replace the expertise-driven process of manual construction with data-driven optimization~\citep{zhang2024aflow,zhao20252flow}.

Recent efforts in this domain can be broadly categorized into two categories based on workflow generation granularity. Task-level approaches generate one shared workflow schema for a specific task domain~\citep{hu2025automated,zhang2025multi,zheng2025mermaidflow}.
Conversely, query-level methods dynamically generate bespoke workflows for individual  inputs~\citep{wang2025scoreflow}.
A growing body of evidence suggests that these automatically generated workflows can achieve superior effectiveness and generalization compared with manually designed counterparts~\citep{shang2024agentsquare,li2024autoflow,yuksekgonul2024textgrad}.

\begin{figure}[t]
  \centering
  \includegraphics[width=0.98\linewidth]{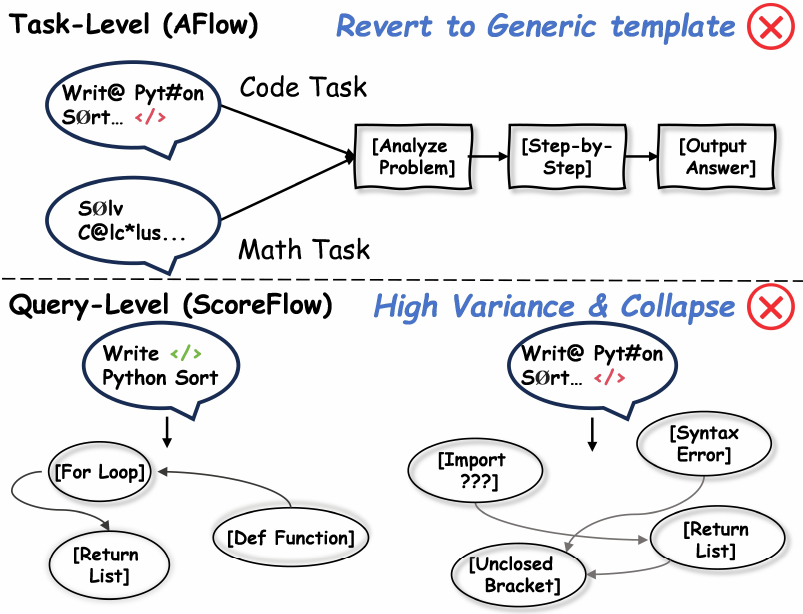}
  \caption{Illustration of two distinct failure modes under perturbation between task-level and query-level methods.}
  \label{fig:difference_query_task}
\end{figure}

However, simply pursuing high performance is insufficient to ensure the stability and reliability of multi-agent systems in real-world deployment.
A critical yet often overlooked attribute is workflow generation robustness. It refers to the ability to generate consistent workflows for semantically equivalent instructions expressed in different forms.
Our evaluation shows that existing methods can achieve robustness scores as low as 0.37 under heavy noise perturbations.
We also observe that task-level and query-level methods exhibit distinct failure modes under input perturbations, as illustrated in Fig.~\ref{fig:difference_query_task} and further analyzed in Section~\ref{ssec:motivation}.
Crucially, prior research~\citep{song2024good,he2025nondeterminism,atil2024non} shows that this instability persists even when sampling temperature of LLMs is reduced to zero. This indicates that the problem is not merely an artifact of randomness, but rather a fundamental failure to achieve semantic invariance.

Motivated by the observations above, we systematically study the robustness problem in existing agentic workflow generation methods.
To enable controlled evaluation, we construct a comprehensive benchmark from 1,255 base tasks spanning 6
benchmarks and 3 domains. For each task, we generate 5 semantically equivalent instruction variants through requirement augmentation, paraphrasing, and noise injection, yielding an extensive testbed of 7,530 queries and 31,889 workflows. 
Through experiments on this benchmark, the robustness problem can be quantitatively verified. When provided with semantically equivalent but differently phrased queries, 
robustness of workflow generation methods
drops from approximately 70\% to as low as 40\%. 
To address this issue, we propose RobustFlow, a framework that improves structural robustness by learning a shared canonical workflow for each group of semantically equivalent queries. RobustFlow selects the workflow that jointly exhibits the highest execution quality and occurrence frequency as the canonical positive sample. Optimizing toward this canonical structure aligns different query formulations with a unified workflow, thereby reducing structural variations induced by query perturbations.
Our contributions are fourfold:
\begin{itemize}
    \item We conduct a systematic investigation into the robustness of agentic workflow generation methods. Our empirical analysis is the first to reveal distinct robustness features of query-level and task-level generation methods, identifying their failure modes under noise perturbations.
    \item We propose a metric framework to quantitatively measure workflow generation robustness by integrating node-level alignment and graph-level similarity.
    \item We construct a comprehensive benchmark dataset comprising 31,889 workflows derived from 1,255 base queries across 6 benchmarks spanning 3 domains. For each base task description, we generate 5 types of semantically equivalent perturbed query variations.
    \item We propose RobustFlow, a training framework that effectively mitigates the identified robustness issues. By leveraging Preference Optimization to align outputs with canonical structures, RobustFlow significantly boosts robustness scores to 70\% - 90\%.
\end{itemize}

\section{Related Work}
\subsection{Automated Agentic Workflow Generation}

Agentic workflow generation aims to automatically construct executable workflows that coordinate multiple agent or operator invocations for a given task~\citep{chen2023autoagents}.
Formally, given a query space $\mathcal Q$ and a set of callable components $\mathcal A$, a workflow generator $G_\theta$ maps a query $q\in\mathcal Q$ to an executable workflow $w=G_\theta(q;\mathcal A)\in\mathcal W$, where $\mathcal W$ denotes workflow search space and $\theta$ denotes trainable parameters of generator. We represent each workflow as a normalized graph $\Gamma(w)=(V,E)$, where nodes $v\in V$ denote component invocations and directed edges $e\in E$ encode data and control dependencies. Based on generation granularity, existing approaches can be categorized as task-level and query-level.

Task-level methods generate one workflow $w$ for each task family $\{q^{(i)}\}_{i=1}^{n}\subseteq\mathcal Q$.
ADAS employs meta-agent search over code-represented workflows~\citep{hu2025automated}, AFlow applies MCTS over operator graphs~\citep{zhang2024aflow}, MaAS searches an agentic supernet to balance performance and cost~\citep{zhang2025multi}, and MermaidFlow performs constrained evolutionary search in a declarative graph language~\citep{zheng2025mermaidflow}. 
In contrast, query-level methods generate one workflow $w$ for each query $q^{(i)}$.
FlowReasoner uses a reasoning-oriented meta-agent~\citep{gao2025flowreasoner}, ScoreFlow applies score-based DPO~\citep{wang2025scoreflow}, and Flow dynamically updates activity-on-vertex graphs based on execution history~\citep{niu2025flow}.

\subsection{Robustness of LLMs} 
The sensitivity of LLMs to semantically equivalent prompt variations has motivated a range of robustness-enhancing approaches at both training and inference time~\citep{wang2020infobert,zhu2023promptrobust}.
Training-time methods improve invariance by exposing models to adversarially perturbed instructions~\citep{goodfellow2014explaining}, aligning representations of semantically equivalent instructions while separating non-equivalent ones~\citep{yan2024contrastive}, or minimizing worst-case consistency risk across instruction variants through distributionally robust preference optimization~\citep{zhao2024improving,fisch2024robust}. At inference time, learned rewriters canonicalize user instructions~\citep{fu2024learning}, while multi-view methods aggregate outputs generated from multiple paraphrases~\citep{chen2024self,yao2024large}.

\section{Preliminary}

\begin{figure*}[t]
  \centering
  \includegraphics[width=0.95\linewidth]{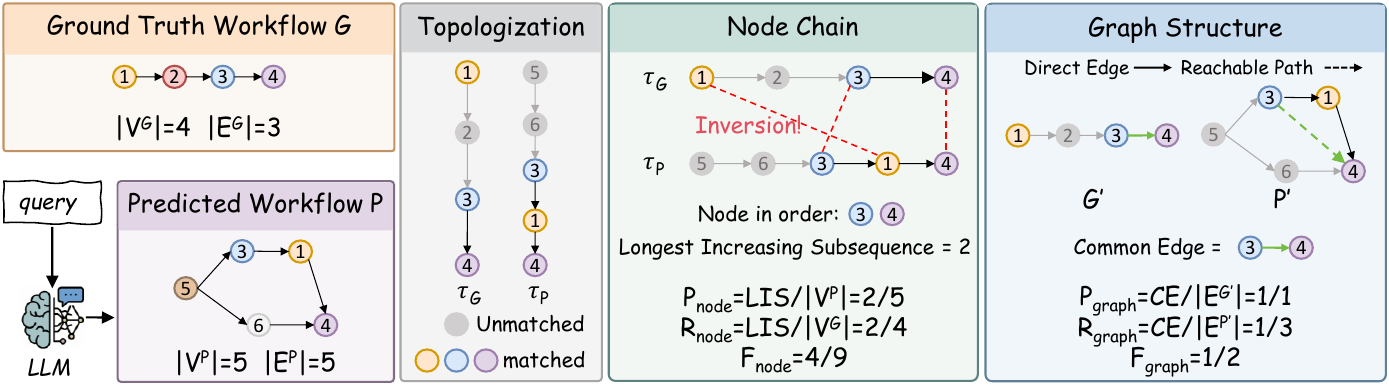}
  \caption{Structure-aware robustness evaluation metrics. We first align nodes between the ground truth and predicted workflows to establish semantic correspondence. Based on this alignment, we compute two key metrics: Node-Chain Robustness, derived from the longest increasing subsequence ($LIS$) of the matched nodes topological order; and Graph-Structure Robustness, calculated by evaluating the preservation of reachable dependency edges in the aligned DAGs.}
  \label{fig:evaluation}
\end{figure*}


In this section, we establish a systematic evaluation framework for workflow robustness by constructing a perturbation benchmark, defining structure-aware metrics, and empirically analyzing the failure modes of existing methods.

\subsection{Perturbation Protocol and Dataset}
\label{ssec:perturbation}

We consider 3 types of input perturbation. \texttt{Paraphrasing} 
modify wording and syntax while preserving the complete task specification. \texttt{Requirement augmentation} introduce non-contradictory constraints, such as resource or output-format requirements, while preserving the core task objective and feasibility. Following TextAttack~\citep{morris2020textattack}, \texttt{Noise injection} applies random word-level substitution, insertion, deletion, and swap. We define light, moderate, and heavy noise using perturbation rates $\rho\in[0.2,0.4]$, $[0.4,0.6]$, and $[0.6,0.8]$ respectively, while protecting task-critical spans. Detailed generation prompts and examples are provided in supplementary material B.

Starting from 1,255 base instructions collected from 6 benchmarks across 3 domains, we generate 5 semantically equivalent variants for each instruction through requirement augmentation, paraphrasing, and noise injection. For each base instruction $q\in\mathcal Q$, we construct an instruction cluster
$\mathcal C(q)=\{q,q^{(1)},\ldots,q^{(5)}\}$,
which contains the original instruction and its 5 variants. This process yields 1,255 instruction clusters comprising 7,530 instructions and 31,889 workflows. All clusters were manually verified by 5 authors over 2 months to ensure that the variants preserve the core task objective, contain feasible and non-contradictory constraints, and maintain valid formatting.

\subsection{Structure-Aware Robustness Evaluation}
\label{ssec:structure}
For a semantic cluster $\mathcal C(q)$, we use the workflow generated from the original instruction $q$ as the ground truth workflow $w^G$, and the workflow generated from a variant $q^{(i)}\in\mathcal C(q)$ as the predicted workflow $w^P$. Their normalized graph representations are denoted by $\Gamma(w^G)=(V^G,E^G)$ and $\Gamma(w^P)=(V^P,E^P)$, respectively. As illustrated in Fig.~\ref{fig:evaluation}, our evaluation first aligns semantically corresponding nodes and then measures structural consistency from two complementary perspectives. Node-chain robustness $\mathcal F_{\mathrm{node}}$ evaluates whether workflow steps and their relative execution order are preserved, whereas graph-structure robustness $\mathcal F_{\mathrm{graph}}$ evaluates whether dependencies among aligned steps remain consistent. We report $\mathbf S(w^G,w^P)=\left(\mathcal F_{\mathrm{node}},\mathcal F_{\mathrm{graph}}\right)$ as the robustness score, where both scores lie in $[0,1]$.




\textbf{Node Alignment.}
We align ground truth nodes $V^G$ with predicted nodes $V^P$ by computing pairwise cosine similarities between their text-based embeddings and discarding pairs below threshold $\beta$. We then solve maximum-weight bipartite matching over the remaining pairs to obtain an optimal one-to-one mapping $\pi: V^{P'}\!\to V^{G'}$. This process identifies the subset of matched nodes (colored nodes in Fig.~\ref{fig:evaluation}) while leaving irrelevant steps (gray nodes) unmatched.

\textbf{Node Chain.}
Following T-eval~\citep{chen2023t} and WorfBench~\citep{qiao2024benchmarking}, we map the aligned nodes in a topological order of $w^P$ to their positions in a topological order of $w^G$, producing an index sequence $\mathbf s$. Its longest increasing subsequence (LIS) length $l$ represents the largest subset of aligned nodes whose relative order is preserved. We define node-chain robustness score $\mathcal F_{\text{node}}$ as:
\begin{equation}
    P_{\text{node}}\!=\!\frac{l}{|V^{P}|},\; R_{\text{node}}\!=\!\frac{l}{|V^{G}|},\;  \mathcal F_{\text{node}}\!=\!\frac{2\,P_{\text{node}}\,R_{\text{node}}}{P_{\text{node}}+R_{\text{node}}}.
\end{equation}
If multiple ground truth topological sequences exist, we select the one maximizing $l$.


\begin{figure}[!t]
  \centering
  \includegraphics[width=0.98\linewidth]{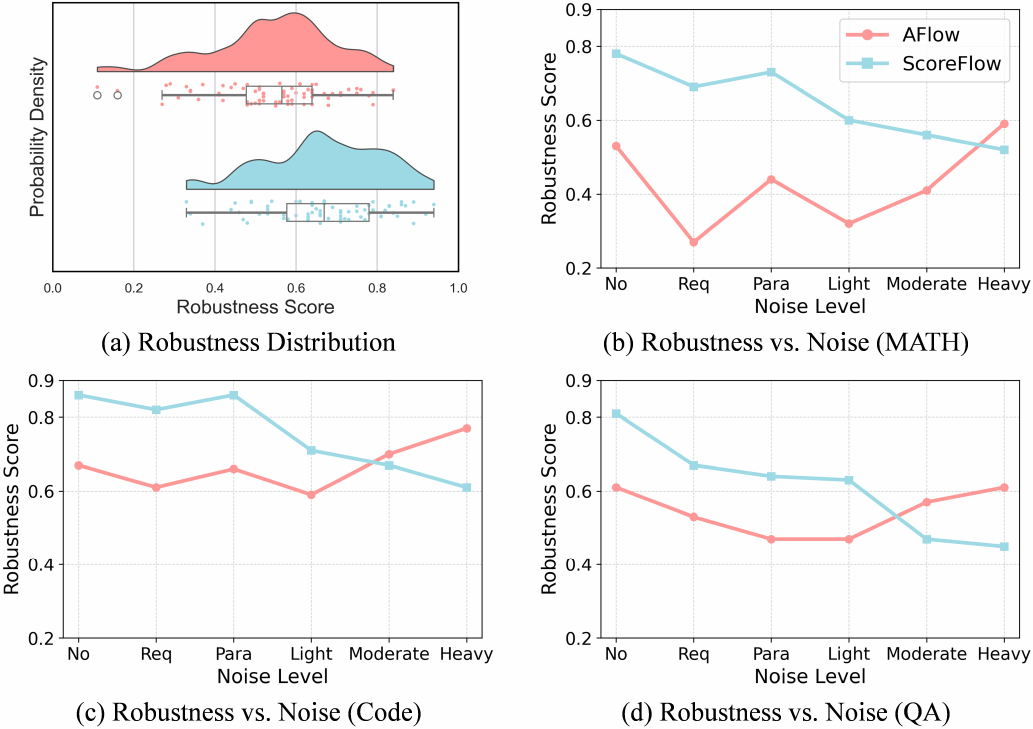}
  \caption{
  Quantitative analysis of workflow structural robustness under varying noise intensities.}
  \label{fig:motivation}
\end{figure}


\begin{figure*}[htbp]
  \centering
  \includegraphics[width=0.98\linewidth]{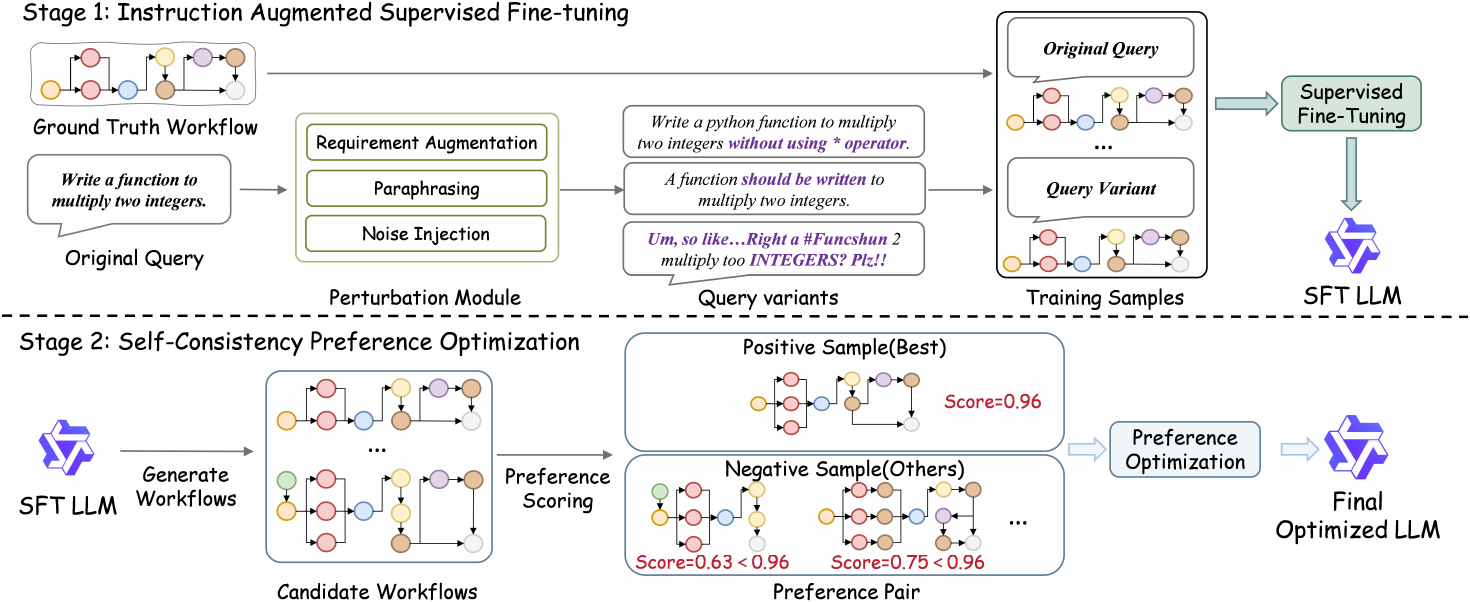}
  \caption{
  Overview of RobustFlow. The training process consists of two stages: (1) We utilize a Perturbation Module to expand the original query into diverse variants and pair these variants with ground truth workflows to fine-tune the Original LLM. (2) We utilize the SFT LLM to generate multiple candidate workflows and evaluate them via preference scoring. Then we select the highest-scoring workflow as the positive sample and lower-scoring ones as negative samples to optimize the final LLM.
  }
  \label{fig:method}
\end{figure*}

\textbf{Graph Structure.}
We compare reachability relations among aligned nodes, allowing a direct dependency in one workflow to be represented by a multi-step path in the other. Let
\begin{equation}
\begin{aligned}
    \mathcal R_G
    &=
    \{(u,v) \mid u\leadsto v \text{ in } \Gamma(w^G)\},\\
    \mathcal R_P
    &=
    \{(\pi(u),\pi(v))
      ,\
      u\leadsto v \text{ in } \Gamma(w^P)\}.
\end{aligned}
\end{equation}
where $\leadsto$ denotes the existence of a path from $u$ to $v$. We define graph-structure robustness score $\mathcal F_{\text{graph}}$ as:
\begin{equation}
\begin{gathered}
    P_{\text{graph}} = \frac{|\mathcal R_{\text{P}} \cap \mathcal R_{\text{G}}|}{|\mathcal R_{\text{P}}|}, \quad 
    R_{\text{graph}} = \frac{|\mathcal R_{\text{P}} \cap \mathcal R_{\text{G}}|}{|\mathcal R_{\text{G}}|}, \\
    \mathcal F_{\text{graph}} = \frac{2P_{\text{graph}}R_{\text{graph}}}{P_{\text{graph}} + R_{\text{graph}}}.
\end{gathered}
\end{equation}

\subsection{Motivational Analysis}
\label{ssec:motivation}
Based on the benchmark and evaluation metrics above, we evaluate two representative agentic workflow generation methods: AFlow, a task-level method, and ScoreFlow, a query-level method. 
As shown in Fig.~\ref{fig:motivation}(a), even without input perturbations, AFlow and ScoreFlow achieve average robustness scores of only 0.60 and 0.81, respectively. Their robustness distributions are also broadly dispersed rather than concentrated near 1.0. 
These results indicate that structural inconsistency is already present under clean inputs and constitutes a shared limitation of the evaluated task-level and query-level methods.






To characterize these failure patterns, we examine robustness across perturbation intensities and 3 task domains in Fig.~\ref{fig:motivation}(b), \ref{fig:motivation}(c), and \ref{fig:motivation}(d). The query-level method ScoreFlow generally degrades as noise increases, suggesting sensitivity to surface-level lexical variations. In contrast, the task-level method AFlow exhibits a non-monotonic trend and may recover under heavy noise, suggesting reduced sensitivity to input-specific details and a tendency to fall back on a generic task-level workflow. As illustrated in Fig.~\ref{fig:difference_query_task}, these contrasting trends indicate distinct degradation patterns in query-level and task-level workflow generation.


\section{RobustFlow}
Motivated by these findings, we propose RobustFlow, a training framework that improves structural robustness by aligning semantically equivalent instruction variants within each cluster to a shared canonical workflow. Unlike existing workflow generators that primarily optimize task performance or efficiency, RobustFlow explicitly targets structural consistency. As shown in Fig.~\ref{fig:method}, RobustFlow follows a two-stage training pipeline. At inference, the resulting generator maps a single query directly to an executable workflow.



\begin{table*}[t]
  \centering
  \small
  \setlength{\tabcolsep}{4pt}
  \begin{tabular}{@{}ll|cccc|cccc|cccc|c|c@{}}
    \toprule
    \multirow{3}{*}{Perturbation}
    & \multirow{3}{*}{Method}
    & \multicolumn{4}{c|}{Code}
    & \multicolumn{4}{c|}{MATH}
    & \multicolumn{4}{c|}{QA}
    & \multirow{3}{*}{Avg.}
    & \multirow{3}{*}{Ratio$\uparrow$} \\

    & & \multicolumn{2}{c}{MBPP}
    & \multicolumn{2}{c|}{HumanEval}
    & \multicolumn{2}{c}{GSM8K}
    & \multicolumn{2}{c|}{MATH}
    & \multicolumn{2}{c}{HotpotQA}
    & \multicolumn{2}{c|}{DROP}
    & & \\

    & & Node & Graph
    & Node & Graph
    & Node & Graph
    & Node & Graph
    & Node & Graph
    & Node & Graph
    & & \\
    \hline

    \multirow{4}{*}{
      \begin{tabular}[l]{@{}l@{}}
        Without\\
        Perturbation
      \end{tabular}
    }
      & AFlow
      & 0.68 & 0.62
      & 0.71 & 0.65
      & 0.58 & 0.52
      & 0.55 & 0.48
      & 0.64 & 0.59
      & 0.62 & 0.57
      & 0.60
      & \multirow{4}{*}{26\%} \\

      & Flow
      & 0.84 & 0.78
      & 0.72 & 0.55
      & - & -
      & - & -
      & - & -
      & - & -
      & 0.72
      & \\

      & ScoreFlow
      & \underline{0.86} & \underline{0.82}
      & \underline{0.89} & \underline{0.85}
      & \underline{0.81} & \underline{0.78}
      & \underline{0.79} & \underline{0.75}
      & \underline{0.83} & \underline{0.79}
      & \underline{0.82} & \underline{0.78}
      & \underline{0.81}
      & \\

      & RobustFlow
      & \textbf{0.99} & \textbf{0.98}
      & \textbf{0.99} & \textbf{0.99}
      & \textbf{0.98} & \textbf{0.97}
      & \textbf{0.96} & \textbf{0.95}
      & \textbf{0.98} & \textbf{0.97}
      & \textbf{0.98} & \textbf{0.96}
      & \textbf{0.97}
      & \\
    \hline

    \multirow{4}{*}{Requirement}
      & AFlow
      & 0.57 & \underline{0.79}
      & 0.62 & 0.47
      & 0.37 & 0.27
      & 0.31 & 0.11
      & 0.56 & 0.50
      & 0.56 & 0.51
      & 0.44
      & \multirow{4}{*}{23\%} \\

      & Flow
      & \textbf{0.91} & \textbf{0.83}
      & 0.73 & 0.42
      & - & -
      & - & -
      & - & -
      & - & -
      & 0.63
      & \\

      & ScoreFlow
      & 0.77 & 0.75
      & \underline{0.89} & \underline{0.86}
      & \underline{0.71} & \underline{0.69}
      & \underline{0.77} & \underline{0.60}
      & \underline{0.66} & \underline{0.72}
      & \underline{0.66} & \underline{0.63}
      & \underline{0.71}
      & \\

      & RobustFlow
      & \underline{0.89} & \textbf{0.83}
      & \textbf{0.95} & \textbf{0.93}
      & \textbf{0.87} & \textbf{0.77}
      & \textbf{0.84} & \textbf{0.80}
      & \textbf{0.88} & \textbf{0.83}
      & \textbf{0.86} & \textbf{0.82}
      & \textbf{0.82}
      & \\
    \hline

    \multirow{4}{*}{Paraphrasing}
      & AFlow
      & 0.51 & \underline{0.75}
      & 0.76 & 0.61
      & 0.41 & 0.34
      & 0.59 & 0.42
      & 0.55 & 0.29
      & 0.51 & 0.52
      & 0.49
      & \multirow{4}{*}{24\%} \\

      & Flow
      & 0.82 & 0.73
      & 0.78 & 0.67
      & - & -
      & - & -
      & - & -
      & - & -
      & 0.70
      & \\

      & ScoreFlow
      & \underline{0.84} & 0.73
      & \underline{0.94} & \underline{0.92}
      & \underline{0.66} & \underline{0.67}
      & \underline{0.83} & \underline{0.76}
      & \underline{0.64} & \underline{0.64}
      & \underline{0.63} & \underline{0.66}
      & \underline{0.73}
      & \\

      & RobustFlow
      & \textbf{0.92} & \textbf{0.88}
      & \textbf{0.98} & \textbf{0.96}
      & \textbf{0.91} & \textbf{0.88}
      & \textbf{0.87} & \textbf{0.83}
      & \textbf{0.93} & \textbf{0.88}
      & \textbf{0.94} & \textbf{0.85}
      & \textbf{0.88}
      & \\
    \hline

    \multirow{4}{*}{Light Noise}
      & AFlow
      & 0.58 & 0.67
      & 0.60 & 0.52
      & 0.42 & 0.36
      & 0.33 & 0.16
      & 0.51 & 0.34
      & 0.52 & 0.49
      & 0.42
      & \multirow{4}{*}{34\%} \\

      & Flow
      & \underline{0.81} & \underline{0.83}
      & 0.67 & 0.45
      & - & -
      & - & -
      & - & -
      & - & -
      & \underline{0.64}
      & \\

      & ScoreFlow
      & 0.54 & 0.71
      & \underline{0.87} & \underline{0.71}
      & \underline{0.64} & \underline{0.53}
      & \underline{0.71} & \underline{0.51}
      & \underline{0.67} & \underline{0.57}
      & \underline{0.71} & \underline{0.57}
      & 0.60
      & \\

      & RobustFlow
      & \textbf{0.95} & \textbf{0.93}
      & \textbf{0.96} & \textbf{0.90}
      & \textbf{0.92} & \textbf{0.85}
      & \textbf{0.88} & \textbf{0.89}
      & \textbf{0.91} & \textbf{0.87}
      & \textbf{0.94} & \textbf{0.89}
      & \textbf{0.89}
      & \\
    \hline

    \multirow{4}{*}{
      \begin{tabular}[l]{@{}l@{}}
        Moderate\\
        Noise
      \end{tabular}
    }
      & AFlow
      & 0.64 & \textbf{0.79}
      & \underline{0.73} & 0.64
      & 0.48 & 0.45
      & 0.41 & 0.28
      & \underline{0.64} & \underline{0.48}
      & 0.58 & \underline{0.57}
      & 0.54
      & \multirow{4}{*}{21\%} \\

      & Flow
      & \underline{0.69} & 0.71
      & 0.53 & 0.59
      & - & -
      & - & -
      & - & -
      & - & -
      & \underline{0.65}
      & \\

      & ScoreFlow
      & 0.63 & 0.67
      & \underline{0.73} & \underline{0.65}
      & \underline{0.61} & \underline{0.51}
      & \underline{0.60} & \underline{0.53}
      & 0.44 & 0.33
      & \underline{0.63} & 0.47
      & 0.53
      & \\

      & RobustFlow
      & \textbf{0.82} & \underline{0.78}
      & \textbf{0.87} & \textbf{0.84}
      & \textbf{0.78} & \textbf{0.71}
      & \textbf{0.83} & \textbf{0.75}
      & \textbf{0.85} & \textbf{0.82}
      & \textbf{0.83} & \textbf{0.79}
      & \textbf{0.78}
      & \\
    \hline

    \multirow{4}{*}{Heavy Noise}
      & AFlow
      & \textbf{0.75} & \textbf{0.84}
      & 0.78 & \underline{0.71}
      & \underline{0.61} & \underline{0.68}
      & \underline{0.63} & 0.44
      & \underline{0.69} & \underline{0.59}
      & 0.50 & \underline{0.64}
      & \underline{0.65}
      & \multirow{4}{*}{22\%} \\

      & Flow
      & 0.64 & 0.40
      & 0.42 & 0.33
      & - & -
      & - & -
      & - & -
      & - & -
      & 0.37
      & \\

      & ScoreFlow
      & 0.48 & 0.46
      & \underline{0.81} & 0.68
      & \underline{0.61} & 0.53
      & 0.49 & \underline{0.45}
      & 0.37 & 0.49
      & \underline{0.58} & 0.34
      & 0.49
      & \\

      & RobustFlow
      & \underline{0.72} & \underline{0.69}
      & \textbf{0.90} & \textbf{0.82}
      & \textbf{0.68} & \textbf{0.72}
      & \textbf{0.78} & \textbf{0.66}
      & \textbf{0.78} & \textbf{0.75}
      & \textbf{0.78} & \textbf{0.71}
      & \textbf{0.72}
      & \\
    \bottomrule
  \end{tabular}

  \caption{
    Comparison of robustness performance among agentic workflow generation
    methods on Code, Math, and QA benchmarks under five distinct and
    challenging perturbation types. We evaluate both node-level and graph-level
    robustness and report the average metrics obtained over ten independent
    runs.
  }
  \label{tab:robustness-evaluation}
\end{table*}

\subsection{Problem Definition}
\label{ssec:problem}



Let workflow generator $G_\theta$ map a natural-language query $q$ to an executable workflow $w$. Given the semantic cluster $\mathcal C(q)$, structural robustness requires $G_\theta(q^{(i)})$, the workflows generated from semantically equivalent variants to remain consistent with $G_\theta(q)$ for every variant $q^{(i)}$. 
Using the structure-aware similarity $\mathcal F$ derived from $\mathcal F_{\text{node}}$ and $\mathcal F_{\text{graph}}$, we define robustness risk $\mathcal R(G_\theta)$ as the expected structural discrepancy over task distribution $\mathcal Q$ and perturbations:
\begin{equation}
    R(G_\theta) = \mathbb{E}_{q\sim\mathcal Q} \left[1-\mathcal S\bigl(G_\theta(q), G_\theta(q^{(i)} \bigr) \right].
\end{equation}
RobustFlow seeks $\theta^\star=\arg\min_\theta R(G_\theta)$ without substantially degrading workflow execution quality.


\subsection{Instruction Augmented Supervised Fine-Tuning}
Instruction augmented supervised fine-tuning (IA-SFT) initializes the generator with executable workflow generation capability and a structural-invariance prior. We use FLORA-Bench~\citep{zhang2025gnns} as base instruction-workflow dataset, denoted by $\mathcal D_0=\{(q_n,\,w_n^{g})\}_{n=1}^{N_0}$, where $q_n$ is the original instruction and $w_n^{g}$ is the ground truth workflow.

Following the perturbation protocol, we generate $K$ semantically preserving variants $q_n^{(i)}$ for each instruction, and let $q_n^{(0)}=q_n$ denote the original. With all variants sharing the same ground truth workflow, we define SFT dataset as:
\begin{equation}
    \mathcal D_{\text{SFT}} =\bigcup_{n=1}^{N_0}\ \big\{\,\big(q_n^{(i)},\,w_n^{g}\big)\ \big|\ i=0,\dots,K\big\},
\end{equation}
where $N_0$ denotes both the number of base instructions and the
resulting semantic clusters. Treating $w_n^g$ as a token sequence $y=(y_1,\dots,y_T)$, we optimize
\begin{equation}   
\mathcal{L}_{\mathrm{SFT}} =\mathbb{E}_{(q,y)\sim \mathcal D_{\text{SFT}}}\Big[-\sum_{t=1}^{T}\log P_{\theta}\big(y_{t}\mid q,\; y_{<t}\big)\Big],
\end{equation}
where $y_{t}$ denotes the $t$-th token, and $y_{<t}$ denotes prefix. We use resulting checkpoint as the initial model $M_0$ for ScPO.

\subsection{Self-Consistency Preference Optimization}

IA-SFT relies on fixed workflow supervision and therefore cannot exploit
structural agreement that emerges among model-generated workflows.
ScPO addresses this limitation by converting intra-cluster
self-consistency into preference supervision while preserving execution
quality.

At iteration $t$, the current model $M_t$ samples $r$ workflows for each query in $\mathcal C(q)$ and aggregates them into a cluster-level candidate multiset $\mathcal Y_q$. Each workflow $w$ is canonicalized as a normalized DAG $\Gamma(w)$. Candidates with identical DAGs are deduplicated, while their multiplicity is retained as the structural frequency $v_q(w)=\left|\left\{w'\in\mathcal Y_q:\Gamma(w')=\Gamma(w)\right\}\right|$.


For each unique candidate $w$, let $s_n(w)\in[0,1]$ denote its execution
quality aggregated over the semantic cluster. Specifically, the
execution metric is final-answer accuracy for mathematical reasoning,
unit-test pass rate for code generation, and F1 for question answering.
We prioritize execution quality and use structural frequency only as a
tie-breaker:
\begin{equation}
w_q^{+}
=
\operatorname*{arg\,max}_{w\in\mathrm{uniq}(\mathcal Y_q)}
\left(s_q(w)+\lambda\,\frac{v_q(w)}{|\mathcal Y_q|}\right).
\end{equation}
The highest-ranked candidate is selected as the canonical positive workflow, while lower-ranked candidates serve as negatives. We pair the same positive workflow $w_q^{+}$ with each query variant in $\mathcal C_q$ to construct the preference dataset $\mathcal D_{\mathrm{pairs}}$.

Initializing $M_\theta$ from $M_t$, we optimize
\begingroup
\small
\begin{equation}
\begin{aligned}
    \mathcal{L}_{\mathrm{ScPO}}(q) = &-\rho_q \log\sigma \left[ \beta \left( \log \frac{M_\theta(w_q^{+} | q)}{M_t(w_q^{+} | q)} - \log \frac{M_\theta(w_q^{-} | q)}{M_t(w_q^{-} | q)} \right) \right] \\
    & -\alpha \rho_q \frac{1}{|w_q^{+}|} \log M_\theta(w_q^{+} \mid q),
\end{aligned}
\end{equation}
\endgroup
where $\rho_n$ denotes the confidence weight, $\beta$ controls preference strength and $\alpha$ controls an auxiliary length-normalized negative log-likelihood term that maintains the absolute likelihood of the positive workflow. After optimization, we set $M_{t+1}\leftarrow M_\theta$ and regenerate candidates for the next
iteration.

\section{Experiments}

\subsection{Experimental Setup}

\textbf{Datasets.} 
Following Section~\ref{ssec:perturbation}, we construct a benchmark dataset of 7,530 query variants and 31,889 workflows across 3 task domains and 6 benchmarks: math reasoning(GSM8K~\citep{cobbe2021training}, MATH~\citep{hendrycks2021measuring}), question answering(HotpotQA~\citep{yang2018hotpotqa}, DROP~\citep{dua2019drop}), and code generation(MBPP~\citep{austin2021program}, HumanEval~\citep{chen2021evaluating}). 

\noindent
\textbf{Baselines.} 
We compare the performance of six agentic workflow generation methods, including AutoAgents~\citep{chen2023autoagents}, ADAS~\citep{hu2025automated}, AFlow~\citep{zhang2024aflow}, MaAS~\citep{zhang2025multi}, ScoreFlow~\citep{wang2025scoreflow}, and FlowReasoner~\citep{gao2025flowreasoner}.
For robustness evaluation, we only compare the fully open-source AFlow, Flow, and ScoreFlow as they require executing the full pipelines under controlled perturbations.

\noindent
\textbf{Implementation Details.}
We use Qwen3-32B~\citep{yang2025qwen3} as primary generator and additionally evaluate Qwen3-1.7B, 4B, 8B, and Llama-3.1-8B to assess scalability and cross-architecture generalization. All generators are served with vLLM~\citep{kwon2023efficient}, and GPT-4o-mini~\citep{hurst2024gpt} serves as the executor. We set all temperatures to 0 and fine-tune with LoRA~\citep{hu2022lora} using ms-swift~\citep{zhao2025swift} on servers with 8 NVIDIA H100 80GB GPUs. Further details of datasets and implementation are provided in supplementary material A.3.




\subsection{Results and Analysis}
We present the main results here, with additional results provided in supplementary material A.

\noindent
\textbf{Overall Robustness.}
As shown in Table~\ref{tab:robustness-evaluation}, structural inconsistency is evident even without input perturbations. AFlow and ScoreFlow achieve average robustness scores of only 0.60 and 0.81, respectively, whereas RobustFlow reaches 0.97. Across the five perturbation settings, RobustFlow maintains average robustness scores between 0.72 and 0.89 and achieves the highest overall average in every setting.

RobustFlow also achieves stronger graph-level robustness across most datasets and perturbation settings, indicating that it better preserves dependencies among workflow steps rather than merely retaining relevant nodes. Although AFlow exhibits a non-monotonic trend and improves under heavy noise, this behavior is consistent with the generic-template fallback identified in our motivational analysis. Overall, RobustFlow preserves task-specific workflow structures more consistently across input variations.

\begin{figure}[!t]
  \centering
  \includegraphics[width=0.98\linewidth]{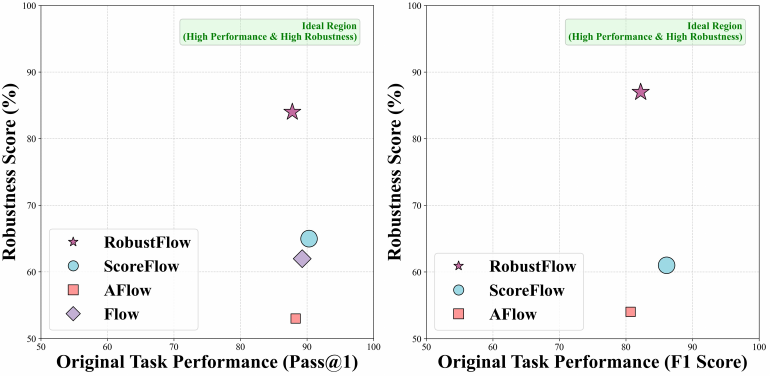}
  \caption{
  Comparison of task performance and robustness across workflow generation methods on Code and QA tasks.}
  \label{fig:pareto}
\end{figure}

\noindent
\textbf{Performance–Robustness Trade-off.}
As shown in Fig.~\ref{fig:pareto}, RobustFlow achieves competitive task performance of 87.79, while remaining below the best-performing query-level baseline (FlowReasoner) by 6.92 points. This result indicates that robustness-oriented optimization introduces a moderate cost in clean-query performance in exchange for substantially improved structural stability.

Qualitative analysis in supplementary meterial suggests that some baseline methods favor compact workflows that perform well on standard evaluations but omit potentially important verification or error-handling steps. In contrast, RobustFlow more consistently preserves complete workflow dependencies across query variations. This rigor prioritizes canonical correctness over fragile optimization, trading slight performance on clean data for defense-in-depth reliability.

\begin{table}[htbp]
  \centering
  \small
  \setlength{\tabcolsep}{5pt}
  \begin{tabular}{@{}lccc@{}}
    \toprule
    Method & \multicolumn{3}{c}{MBPP} \\
    \cmidrule(lr){2-4}
           & Node & Graph & pass@1 \\
    \midrule
    Base generator      & 0.63 & 0.58 & 71.28 \\
    +IA-SFT only        & 0.66 & 0.66 & 83.87 \\
    +ScPO only          & 0.76 & 0.79 & 75.09 \\
    IA-SFT+Score-DPO    & 0.49 & 0.53 & \textbf{86.82} \\
    IA-SFT+ScPO         & \textbf{0.88} & \textbf{0.85} & 81.90 \\
    \bottomrule
  \end{tabular}
  \caption{Ablation study of main components of RobustFlow.}
  \label{tab:component-ablation}
\end{table}

\noindent
\textbf{Ablation Study.}
Table~\ref{tab:component-ablation} examines the contributions of IA-SFT and ScPO on MBPP. Compared with the base model, IA-SFT primarily improves execution performance, whereas ScPO produces larger gains in node- and graph-level robustness. Combining both components achieves the strongest overall robustness while maintaining a pass@1 score of 81.90, confirming their complementary roles in preserving task competence and structural consistency. Under the same IA-SFT initialization, replacing ScPO with Score-DPO improves pass@1 but substantially reduces robustness, indicating that execution-score-based preferences alone do not effectively preserve workflow structure.

\begin{figure*}[htbp]
  \centering
  \includegraphics[width=0.98\linewidth]{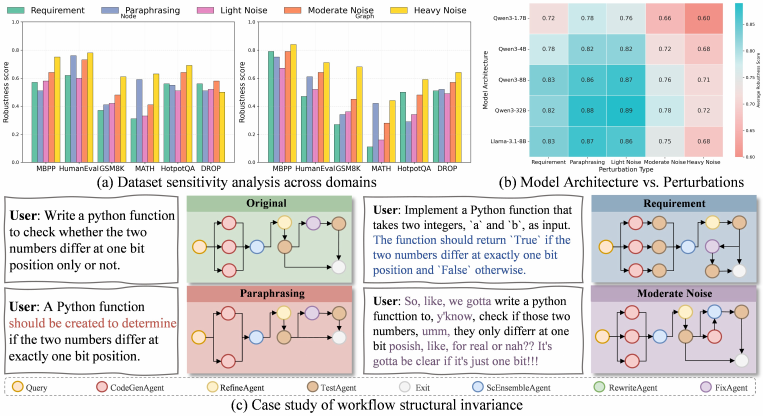}
  \caption{Comprehensive experimental analysis of RobustFlow.}
  \label{fig:experiments}
\end{figure*}

\noindent
\textbf{Architecture and Scale Generalization} 
As shown in Fig.~\ref{fig:experiments}(b), RobustFlow consistently improves structural robustness across Qwen3 and Llama-3.1 model families, covering parameter scales from 1.7B to 32B. These consistent improvements indicate that RobustFlow generalizes across different backbone architectures and model sizes, rather than relying on a specific model family or parameter scale.

\begin{table}[tb]
  \centering
  \small
  \setlength{\tabcolsep}{3pt}
  \begin{tabular}{@{}ll|cccc|c@{}}
    \toprule
    \multirow{2}{*}{Perturbation}
    & \multirow{2}{*}{Method}
    & \multicolumn{2}{c}{HotpotQA}
    & \multicolumn{2}{c|}{DROP}
    & \multirow{2}{*}{Avg.} \\
    & & Node & Graph & Node & Graph & \\
    \hline

    \multirow{3}{*}{
      \begin{tabular}[c]{@{}l@{}}
        Adversarial\\
        Paraphrases
      \end{tabular}
    }
      & AFlow
      & 0.49 & 0.41
      & 0.43 & 0.41
      & 0.43 \\

      & ScoreFlow
      & 0.29 & 0.21
      & 0.26 & 0.23
      & 0.25 \\

      & RobustFlow
      & \textbf{0.83} & \textbf{0.76}
      & \textbf{0.81} & \textbf{0.77}
      & \textbf{0.79} \\
    \hline

    \multirow{3}{*}{
      \begin{tabular}[c]{@{}l@{}}
        Format\\
        Perturbations
      \end{tabular}
    }
      & AFlow
      & 0.57 & 0.49
      & 0.54 & 0.49
      & 0.52 \\

      & ScoreFlow
      & 0.78 & 0.67
      & 0.68 & 0.61
      & 0.68 \\

      & RobustFlow
      & \textbf{0.97} & \textbf{0.94}
      & \textbf{0.96} & \textbf{0.95}
      & \textbf{0.96} \\
    \bottomrule
  \end{tabular}

  \caption{
    Robustness comparison on QA tasks under unseen adversarial paraphrases and format perturbations.
  }
  \label{tab:qa-robustness-detail}
\end{table}


\noindent
\textbf{Generalization to Unseen Perturbations.} 
Table~\ref{tab:qa-robustness-detail} evaluates RobustFlow on QA tasks under two perturbation types not observed during training: adversarial paraphrasing and format perturbations. Adversarial paraphrasing introduces intentionally difficult, semantics-preserving syntactic variations, whereas format perturbations alter input presentation without changing the task content. RobustFlow maintains approximately 80\% structural robustness under adversarial paraphrasing and nearly 95\% under format perturbations, outperforming the baselines in both settings. These results demonstrate that RobustFlow generalizes effectively to unseen perturbation types on QA tasks.

\begin{table}[!t]
  \centering
  \scriptsize
  \setlength{\tabcolsep}{3pt}
  \renewcommand{\arraystretch}{0.95}
  \begin{tabular}{@{}lcccccc@{}}
    \hline
    \multirow{2}{*}{Method}
    & \multicolumn{2}{c}{Training}
    & \multicolumn{4}{c}{Inference / Query} \\
    & Params(B)
    & GPU(hr)
    & Gen.\ Tok.
    & Gen.\ Lat.
    & Exec.\ Tok.
    & Exec.\ Lat. \\
    \hline

    AFlow
    & N/A & 0
    & 50316 & 1394
    & 9568 & 242 \\

    Flow
    & N/A & 0
    & 26097 & 2432
    & 4632 & 96 \\

    ScoreFlow
    & 8 & 268
    & 5883 & 47.45
    & 3199 & 108 \\

    \hline

    \multirow{4}{*}{RobustFlow}
    & 32 & 1327
    & \multirow{4}{*}{1434}
    & 48.11
    & \multirow{4}{*}{6035}
    & \multirow{4}{*}{216} \\

    & 8 & 473
    &
    & 21.10
    &
    & \\

    & 4 & 227
    &
    & 20.56
    &
    & \\

    & 1.7 & 121
    &
    & 16.07
    &
    & \\

    \hline
  \end{tabular}

  \caption{
    Training and per-query inference costs. Gen./Exec. denote workflow generation/execution, Tok. and Lat. denote tokens/latency in seconds.
  }
  \label{tab:cost-comparison}
\end{table}









\noindent
\textbf{Domain-wise Robustness Analysis.}
Figure~\ref{fig:experiments}(a) reveals clear and consistent differences in workflow robustness across task domains. Among the evaluated domains, code generation achieves the highest node- and graph-level robustness, whereas QA exhibits substantially lower robustness, particularly under surface-level noise perturbations. This pattern suggests that tasks with greater linguistic variability and less explicit structural constraints are more challenging for workflow generators to handle consistently.

\noindent
\textbf{Training Cost and Inference Efficiency.}
As shown in Table~\ref{tab:cost-comparison}, RobustFlow shifts computational cost from online workflow search to offline model training. Although RobustFlow-32B incurs higher offline training costs than search-free baselines, it requires only 1,434 generation tokens per query and achieves generation latency comparable to ScoreFlow. Smaller RobustFlow variants further reduce both offline training time and online generation latency. Nevertheless, workflow execution remains the dominant component of total online latency, limiting the end-to-end speedup obtained from faster workflow generation.

\noindent
\textbf{Case Study.} 
As shown in Fig.~\ref{fig:experiments}(c), RobustFlow generates a consistent six-stage workflow for the same task under different perturbations:
\texttt{CodeGenAgent} $\rightarrow$ \texttt{ScEnsembleAgent} $\rightarrow$ \texttt{RefineAgent} $\rightarrow$ \texttt{TestAgent} $\rightarrow$ \texttt{FixAgent} $\rightarrow$ \texttt{Exit}.
Despite minor variations in workflow descriptions, the global topology and control dependencies remain consistent across all perturbations shown.

\section{Conclusion}
In this paper, we systematically investigate structural robustness in agentic workflow generation, an important yet underexplored requirement for reliable deployment across diverse real-world settings. We develop a structure-aware benchmark suite with node- and graph-level metrics, covering 1,255 semantic clusters, 7,530 perturbed queries, and 31,889 generated workflows. We further propose RobustFlow, which combines instruction-augmented SFT with self-consistency preference optimization within semantic clusters. Across datasets and perturbation types, RobustFlow raises average structural robustness to approximately 80\% while maintaining competitive task performance, with only a moderate gap from the strongest query-level baseline. These findings establish structural robustness as an important optimization objective for workflow generators. Future work will jointly optimize robustness, task success, and execution cost, and evaluate generalization across broader domains and tool ecosystems.

\bibliography{main}

@article{chen2023autoagents,
  title={Autoagents: A framework for automatic agent generation},
  author={Chen, Guangyao and Dong, Siwei and Shu, Yu and Zhang, Ge and Sesay, Jaward and Karlsson, B{\"o}rje F and Fu, Jie and Shi, Yemin},
  journal={arXiv preprint arXiv:2309.17288},
  year={2023}
}

@inproceedings{hong2024metagpt,
  title={MetaGPT: Meta programming for a multi-agent collaborative framework},
  author={Hong, Sirui and Zhuge, Mingchen and Chen, Jonathan and Zheng, Xiawu and Cheng, Yuheng and Zhang, Ceyao and Wang, Jinlin and Wang, Zili and Yau, Steven Ka Shing and Lin, Zijuan and others},
  year={2024},
  organization={International Conference on Learning Representations, ICLR}
}

@article{qian2023chatdev,
  title={Chatdev: Communicative agents for software development},
  author={Qian, Chen and Liu, Wei and Liu, Hongzhang and Chen, Nuo and Dang, Yufan and Li, Jiahao and Yang, Cheng and Chen, Weize and Su, Yusheng and Cong, Xin and others},
  journal={arXiv preprint arXiv:2307.07924},
  year={2023}
}

@article{gao2025flowreasoner,
  title={Flowreasoner: Reinforcing query-level meta-agents},
  author={Gao, Hongcheng and Liu, Yue and He, Yufei and Dou, Longxu and Du, Chao and Deng, Zhijie and Hooi, Bryan and Lin, Min and Pang, Tianyu},
  journal={arXiv preprint arXiv:2504.15257},
  year={2025}
}

@inproceedings{hu2025automated,
  title={Automated design of agentic systems},
  author={Hu, Shengran and Lu, Cong and Clune, Jeff},
  booktitle={International Conference on Learning Representations},
  volume={2025},
  pages={21344--21377},
  year={2025}
}

@article{zhang2024aflow,
  title={Aflow: Automating agentic workflow generation},
  author={Zhang, Jiayi and Xiang, Jinyu and Yu, Zhaoyang and Teng, Fengwei and Chen, Xionghui and Chen, Jiaqi and Zhuge, Mingchen and Cheng, Xin and Hong, Sirui and Wang, Jinlin and others},
  journal={arXiv preprint arXiv:2410.10762},
  year={2024}
}

@article{zhang2025multi,
  title={Multi-agent architecture search via agentic supernet},
  author={Zhang, Guibin and Niu, Luyang and Fang, Junfeng and Wang, Kun and Bai, Lei and Wang, Xiang},
  journal={arXiv preprint arXiv:2502.04180},
  year={2025}
}

@article{wang2025scoreflow,
  title={Scoreflow: Mastering llm agent workflows via score-based preference optimization},
  author={Wang, Yinjie and Yang, Ling and Li, Guohao and Wang, Mengdi and Aragam, Bryon},
  journal={arXiv preprint arXiv:2502.04306},
  year={2025}
}

@article{niu2025flow,
  title={Flow: Modularized agentic workflow automation},
  author={Niu, Boye and Song, Yiliao and Lian, Kai and Shen, Yifan and Yao, Yu and Zhang, Kun and Liu, Tongliang},
  journal={arXiv preprint arXiv:2501.07834},
  year={2025}
}

@article{zhang2025gnns,
  title={Gnns as predictors of agentic workflow performances},
  author={Zhang, Yuanshuo and Hou, Yuchen and Tang, Bohan and Chen, Shuo and Zhang, Muhan and Dong, Xiaowen and Chen, Siheng},
  journal={arXiv preprint arXiv:2503.11301},
  year={2025}
}

@article{wang2020infobert,
  title={Infobert: Improving robustness of language models from an information theoretic perspective},
  author={Wang, Boxin and Wang, Shuohang and Cheng, Yu and Gan, Zhe and Jia, Ruoxi and Li, Bo and Liu, Jingjing},
  journal={arXiv preprint arXiv:2010.02329},
  year={2020}
}

@inproceedings{zhu2023promptrobust,
  title={Promptrobust: Towards evaluating the robustness of large language models on adversarial prompts},
  author={Zhu, Kaijie and Wang, Jindong and Zhou, Jiaheng and Wang, Zichen and Chen, Hao and Wang, Yidong and Yang, Linyi and Ye, Wei and Zhang, Yue and Gong, Neil and others},
  booktitle={Proceedings of the 1st ACM workshop on large AI systems and models with privacy and safety analysis},
  pages={57--68},
  year={2023}
}

@article{goodfellow2014explaining,
  title={Explaining and harnessing adversarial examples},
  author={Goodfellow, Ian J and Shlens, Jonathon and Szegedy, Christian},
  journal={arXiv preprint arXiv:1412.6572},
  year={2014}
}

@article{yan2024contrastive,
  title={Contrastive instruction tuning},
  author={Yan, Tianyi Lorena and Wang, Fei and Huang, James Y and Zhou, Wenxuan and Yin, Fan and Galstyan, Aram and Yin, Wenpeng and Chen, Muhao},
  journal={arXiv preprint arXiv:2402.11138},
  year={2024}
}

@article{zhao2024improving,
  title={Improving the robustness of large language models via consistency alignment},
  author={Zhao, Yukun and Yan, Lingyong and Sun, Weiwei and Xing, Guoliang and Wang, Shuaiqiang and Meng, Chong and Cheng, Zhicong and Ren, Zhaochun and Yin, Dawei},
  journal={arXiv preprint arXiv:2403.14221},
  year={2024}
}

@article{fisch2024robust,
  title={Robust preference optimization through reward model distillation},
  author={Fisch, Adam and Eisenstein, Jacob and Zayats, Vicky and Agarwal, Alekh and Beirami, Ahmad and Nagpal, Chirag and Shaw, Pete and Berant, Jonathan},
  journal={arXiv preprint arXiv:2405.19316},
  year={2024}
}

@inproceedings{chen2024self,
  title={Self-Para-Consistency: Improving Reasoning Tasks at Low Cost for Large Language Models},
  author={Chen, Wenqing and Wang, Weicheng and Chu, Zhixuan and Ren, Kui and Zheng, Zibin and Lu, Zhichao},
  booktitle={62nd Annual Meeting of the Association for Computational Linguistics (ACL 2024)},
  pages={14162--14167},
  year={2024},
  organization={Association for Computational Linguistics}
}

@article{yao2024large,
  title={Large language models are contrastive reasoners},
  author={Yao, Liang},
  journal={arXiv preprint arXiv:2403.08211},
  year={2024}
}

@inproceedings{fu2024learning,
  title={Learning to paraphrase for alignment with LLM preference},
  author={Fu, Junbo and Zhao, Guoshuai and Deng, Yimin and Mi, Yunqi and Qian, Xueming},
  booktitle={Findings of the Association for Computational Linguistics: EMNLP 2024},
  pages={2394--2407},
  year={2024}
}

@article{austin2021program,
  title={Program synthesis with large language models},
  author={Austin, Jacob and Odena, Augustus and Nye, Maxwell and Bosma, Maarten and Michalewski, Henryk and Dohan, David and Jiang, Ellen and Cai, Carrie and Terry, Michael and Le, Quoc and others},
  journal={arXiv preprint arXiv:2108.07732},
  year={2021}
}

@article{cobbe2021training,
  title={Training verifiers to solve math word problems},
  author={Cobbe, Karl and Kosaraju, Vineet and Bavarian, Mohammad and Chen, Mark and Jun, Heewoo and Kaiser, Lukasz and Plappert, Matthias and Tworek, Jerry and Hilton, Jacob and Nakano, Reiichiro and others},
  journal={arXiv preprint arXiv:2110.14168},
  year={2021}
}

@article{hendrycks2021measuring,
  title={Measuring mathematical problem solving with the math dataset},
  author={Hendrycks, Dan and Burns, Collin and Kadavath, Saurav and Arora, Akul and Basart, Steven and Tang, Eric and Song, Dawn and Steinhardt, Jacob},
  journal={arXiv preprint arXiv:2103.03874},
  year={2021}
}

@article{yang2018hotpotqa,
  title={HotpotQA: A dataset for diverse, explainable multi-hop question answering},
  author={Yang, Zhilin and Qi, Peng and Zhang, Saizheng and Bengio, Yoshua and Cohen, William W and Salakhutdinov, Ruslan and Manning, Christopher D},
  journal={arXiv preprint arXiv:1809.09600},
  year={2018}
}

@article{dua2019drop,
  title={DROP: A reading comprehension benchmark requiring discrete reasoning over paragraphs},
  author={Dua, Dheeru and Wang, Yizhong and Dasigi, Pradeep and Stanovsky, Gabriel and Singh, Sameer and Gardner, Matt},
  journal={arXiv preprint arXiv:1903.00161},
  year={2019}
}

@article{yang2025qwen3,
  title={Qwen3 technical report},
  author={Yang, An and Li, Anfeng and Yang, Baosong and Zhang, Beichen and Hui, Binyuan and Zheng, Bo and Yu, Bowen and Gao, Chang and Huang, Chengen and Lv, Chenxu and others},
  journal={arXiv preprint arXiv:2505.09388},
  year={2025}
}

@inproceedings{kwon2023efficient,
  title={Efficient memory management for large language model serving with pagedattention},
  author={Kwon, Woosuk and Li, Zhuohan and Zhuang, Siyuan and Sheng, Ying and Zheng, Lianmin and Yu, Cody Hao and Gonzalez, Joseph and Zhang, Hao and Stoica, Ion},
  booktitle={Proceedings of the 29th symposium on operating systems principles},
  pages={611--626},
  year={2023}
}

@article{hurst2024gpt,
  title={Gpt-4o system card},
  author={Hurst, Aaron and Lerer, Adam and Goucher, Adam P and Perelman, Adam and Ramesh, Aditya and Clark, Aidan and Ostrow, AJ and Welihinda, Akila and Hayes, Alan and Radford, Alec and others},
  journal={arXiv preprint arXiv:2410.21276},
  year={2024}
}

@article{hu2022lora,
  title={Lora: Low-rank adaptation of large language models.},
  author={Hu, Edward J and Shen, Yelong and Wallis, Phillip and Allen-Zhu, Zeyuan and Li, Yuanzhi and Wang, Shean and Wang, Lu and Chen, Weizhu and others},
  journal={ICLR},
  volume={1},
  number={2},
  pages={3},
  year={2022}
}

@inproceedings{zhao2025swift,
  title={Swift: a scalable lightweight infrastructure for fine-tuning},
  author={Zhao, Yuze and Huang, Jintao and Hu, Jinghan and Wang, Xingjun and Mao, Yunlin and Zhang, Daoze and Jiang, Zeyinzi and Wu, Zhikai and Ai, Baole and Wang, Ang and others},
  booktitle={Proceedings of the AAAI Conference on Artificial Intelligence},
  volume={39},
  number={28},
  pages={29733--29735},
  year={2025}
}

@article{morris2020textattack,
  title={Textattack: A framework for adversarial attacks, data augmentation, and adversarial training in nlp},
  author={Morris, John X and Lifland, Eli and Yoo, Jin Yong and Grigsby, Jake and Jin, Di and Qi, Yanjun},
  journal={arXiv preprint arXiv:2005.05909},
  year={2020}
}

@article{chen2023t,
  title={T-eval: Evaluating the tool utilization capability of large language models step by step},
  author={Chen, Zehui and Du, Weihua and Zhang, Wenwei and Liu, Kuikun and Liu, Jiangning and Zheng, Miao and Zhuo, Jingming and Zhang, Songyang and Lin, Dahua and Chen, Kai and others},
  journal={arXiv preprint arXiv:2312.14033},
  year={2023}
}

@article{qiao2024benchmarking,
  title={Benchmarking agentic workflow generation},
  author={Qiao, Shuofei and Fang, Runnan and Qiu, Zhisong and Wang, Xiaobin and Zhang, Ningyu and Jiang, Yong and Xie, Pengjun and Huang, Fei and Chen, Huajun},
  journal={arXiv preprint arXiv:2410.07869},
  year={2024}
}

@article{chen2021evaluating,
  title={Evaluating large language models trained on code},
  author={Chen, Mark and Tworek, Jerry and Jun, Heewoo and Yuan, Qiming and Pinto, Henrique Ponde De Oliveira and Kaplan, Jared and Edwards, Harri and Burda, Yuri and Joseph, Nicholas and Brockman, Greg and others},
  journal={arXiv preprint arXiv:2107.03374},
  year={2021}
}

@inproceedings{khattab2024dspy,
  title={Dspy: Compiling declarative language model calls into state-of-the-art pipelines},
  author={Khattab, Omar and Singhvi, Arnav and Maheshwari, Paridhi and Zhang, Zhiyuan and Santhanam, Keshav and Haq, Saiful and Sharma, Ashutosh and Joshi, Thomas T and Moazam, Hanna and Miller, Heather and others},
  booktitle={The Twelfth International Conference on Learning Representations},
  year={2024}
}

@article{tang2023verifai,
  title={VerifAI: verified generative AI},
  author={Tang, Nan and Yang, Chenyu and Fan, Ju and Cao, Lei and Luo, Yuyu and Halevy, Alon},
  journal={arXiv preprint arXiv:2307.02796},
  year={2023}
}

@article{shang2024agentsquare,
  title={Agentsquare: Automatic llm agent search in modular design space},
  author={Shang, Yu and Li, Yu and Zhao, Keyu and Ma, Likai and Liu, Jiahe and Xu, Fengli and Li, Yong},
  journal={arXiv preprint arXiv:2410.06153},
  year={2024}
}

@article{li2024autoflow,
  title={Autoflow: Automated workflow generation for large language model agents},
  author={Li, Zelong and Xu, Shuyuan and Mei, Kai and Hua, Wenyue and Rama, Balaji and Raheja, Om and Wang, Hao and Zhu, He and Zhang, Yongfeng},
  journal={arXiv preprint arXiv:2407.12821},
  year={2024}
}

@article{yuksekgonul2024textgrad,
  title={Textgrad: Automatic" differentiation" via text},
  author={Yuksekgonul, Mert and Bianchi, Federico and Boen, Joseph and Liu, Sheng and Huang, Zhi and Guestrin, Carlos and Zou, James},
  journal={arXiv preprint arXiv:2406.07496},
  year={2024}
}

@article{he2025nondeterminism,
  author = {Horace He and Thinking Machines Lab},
  title = {Defeating Nondeterminism in LLM Inference},
  journal = {Thinking Machines Lab: Connectionism},
  year = {2025},
  note = {https://thinkingmachines.ai/blog/defeating-nondeterminism-in-llm-inference/},
  doi = {10.64434/tml.20250910}
}

@article{atil2024non,
  title={Non-determinism of" deterministic" llm settings},
  author={Atil, Berk and Aykent, Sarp and Chittams, Alexa and Fu, Lisheng and Passonneau, Rebecca J and Radcliffe, Evan and Rajagopal, Guru Rajan and Sloan, Adam and Tudrej, Tomasz and Ture, Ferhan and others},
  journal={arXiv preprint arXiv:2408.04667},
  year={2024}
}

@article{song2024good,
  title={The good, the bad, and the greedy: Evaluation of llms should not ignore non-determinism},
  author={Song, Yifan and Wang, Guoyin and Li, Sujian and Lin, Bill Yuchen},
  journal={arXiv preprint arXiv:2407.10457},
  year={2024}
}

@article{zheng2025mermaidflow,
  title={MermaidFlow: Redefining Agentic Workflow Generation via Safety-Constrained Evolutionary Programming},
  author={Zheng, Chengqi and Chen, Jianda and Lyu, Yueming and Ng, Wen Zheng Terence and Zhang, Haopeng and Ong, Yew-Soon and Tsang, Ivor and Yin, Haiyan},
  journal={arXiv preprint arXiv:2505.22967},
  year={2025}
}

@misc{zhao20252flow,
      title={$A^2Flow:$ Automating Agentic Workflow Generation via Self-Adaptive Abstraction Operators}, 
      author={Mingming Zhao and Xiaokang Wei and Yuanqi Shao and Kaiwen Zhou and Lin Yang and Siwei Rao and Junhui Zhan and Zhitang Chen},
      year={2025},
      eprint={2511.20693},
      archivePrefix={arXiv},
      primaryClass={cs.AI},
      url={https://arxiv.org/abs/2511.20693}, 
}


\end{document}